\documentclass[12pt]{article}

\usepackage{scicite}

\usepackage{times}

\usepackage{amsmath}
\usepackage{amssymb}
\usepackage{graphicx}
\usepackage{color}
\usepackage{amsfonts}
\usepackage{subfigure}
\usepackage{enumitem}
\usepackage{url}
\usepackage[normalem]{ulem}
\usepackage[capitalize]{cleveref}

\newenvironment{sciabstract}{%
\begin{quote} \bf}
{\end{quote}}

\title{Gender disparities in the dissemination and acquisition of scientific knowledge}

\author
{C. Zappal\`a$^{1,\dag,\ast}$, L. Gallo$^{2,3,\dag}$, J. Bachmann$^{4,2}$, F. Battiston$^2$, F. Karimi$^{4,5,\ast}$\\
\normalsize{$^{1}$Center for Collective Learning, Corvinus Institute for Advanced Studies (CIAS),}\\
\normalsize{Corvinus University, 1093 Budapest, Hungary}\\
\normalsize{$^{2}$Department of Network and Data Science,}\\
\normalsize{Central European University, 1100 Vienna, Austria}\\
\normalsize{$^{3}$ANETI Lab, Corvinus Institute for Advanced Studies (CIAS),}\\
\normalsize{Corvinus University, 1093 Budapest, Hungary}\\
\normalsize{$^{4}$Complexity Science Hub Vienna, A-1080 Vienna, Austria} \\
\normalsize{$^{5}$Graz University of Technology, 8010 Graz, Austria,}\\
\normalsize{$^\dag$These authors contributed equally} \\
\normalsize{$^\ast$To whom correspondence should be addressed;}\\ 
\normalsize{E-mail: chiara-zappala@uni-corvinus.hu; karimi@csh.at}
}

\date{}

\begin{document} 


\maketitle


\begin{sciabstract}
Recent research has challenged the widespread belief that gender inequities in academia would disappear simply by increasing the number of women.
More complex causes might be at play, embodied in the networked structure of scientific collaborations.
Here, we aim to understand the structural inequality between women and men in the dissemination of scientific knowledge.
We use a large-scale dataset of academic publications from the American Physical Society (APS) to build a time-varying network of collaborations from 1970 to 2020.
We model knowledge dissemination as a simple contagion process in which scientists become informed based on the propagation of knowledge through their collaborators. 
We quantify the fairness of the system in terms of how women acquire and diffuse knowledge compared to men.
Our results indicate that knowledge acquisition and diffusion are slower for women than expected.
We find that the main determinant of women's disadvantage is the gap in the cumulative number of collaborators, highlighting how time creates structural disadvantages that contribute to marginalizing women in physics.
Our work sheds light on how the dynamics of scientific collaborations shape gender disparities in knowledge dissemination and calls for a deeper understanding of how to intervene to improve fairness and diversity in the scientific community.
\end{sciabstract}


\section*{Introduction}

Gender disparities in academia persist despite growing participation of women. 
Research suggests that women are still disadvantaged on multiple dimensions, such as productivity, citations, salary, funding, hiring, online representation, credit, and recognition \cite{ley2008gender,way2016gender,holman2018gender,charlesworth2019gender,witteman2019gender,huang2020historical,vasarhelyi2021gender,ross2022women,teich2022citation}.
Most evidently, the gap in the number of men and women in Science, Technology, Engineering, and Mathematics (STEM) is indisputable.
In physics, the percentage of women is as low as 15\%, and the increase in women's participation in recent years is still considerably slow \cite{huang2020historical,holman2018gender}.
Combined with the historical barriers to entry into science and the first-mover advantages for men, this gap will not close within this century \cite{holman2018gender,kong2022influence}. 
However, past research has questioned the common belief that increasing the number of female researchers is enough to achieve gender equality in academia \cite{huang2020historical,neuhauser2023improving}.
One reason is that temporal discrepancies in the arrival and participation of women in STEM have far-reaching implications, fueling the persistence of gender stereotypes associated with social roles in STEM which prevents engagement of future generations \cite{carli2016stereotypes} and ultimately impacts hiring and promotion decisions \cite{moss2012science,clifton2019mathematical}.

In this context, the question of how disparities between male and female scholars affect the dissemination of scientific knowledge remains unanswered.
This question is crucial for two reasons: First, gender diversity is fundamental to spark scientific innovation, because it can broaden the perspectives of researchers \cite{page2007,woolley2010evidence,nielsen2017gender};
also, science requires that innovative ideas are effectively communicated to others in order to progress \cite{merton1968matthew}.
These new ideas, and knowledge in general, are not communicated exclusively in written form; rather, they thrive through social interactions.
That is, knowledge is not ``in the air'', i.e., globally available \cite{cowan2005network}, but it is constrained by geographical and social proximity, as hinted by an increasing body of literature \cite{jaffe1993geographic,valente1995network,breschi2003mobility,breschi2009mobility,wuestman2019geography,abramo2020role,duede2024being}.
The reason is that acquiring complex scientific and technical knowledge is a trial-and-error process that requires effort.
Therefore, a close connection to the source of knowledge is crucial for its deep understanding, transmission, and ultimately its use to advance science \cite{sorenson2006complexity}.

In academia, the coauthorship network embodies meaningful social connections that let scientific knowledge circulate.
This network is the expression of the fruitful collaborations authors had during their careers, sharing expertise, skills, and experience \cite{singh2005collaborative,sorenson2006complexity,sekara2018chaperone,venturini2024collaboration}.
Thus, the structure and evolution of collaboration ties between scientists \cite{newman2001structure,newman2001clustering,guimera2005team,sun2013social} 
deeply influence the dissemination of scientific and technical knowledge \cite{espinosa2024co}.

Male and female scientists substantially differ in the way they build their collaboration network throughout their careers.
Women have fewer distinct co-authors, tend to collaborate more with other women forming isolated communities, are less likely to act as academic brokers, and engage in less favorable collaborations \cite{bozeman2004scientists,abramo2013gender,zeng2016differences,jadidi2018gender,li.etal_untanglingnetworkeffects_2022,bachmann2024cumulativeadvantagebrokerageacademia,bunker2024structure}.
Overall, discrepancies in collaboration patterns between men and women can potentially generate gender differences in the flow of knowledge. 

Our hypothesis is that the time-varying collaboration network functions as a structural barrier by limiting the opportunity for women to acquire and diffuse knowledge.
To this end, we use a large-scale dataset of academic publications of the American Physical Society (APS) and build a temporal network \cite{holme2012temporal,masuda2016guide} of scientific collaborations from 1970 to 2020.
Specifically, the network consists of yearly snapshots where nodes are authors and links connect researchers who coauthored a paper during that year.
We model knowledge flow as a simple contagion process on this collaboration network \cite{goffman1966,pastor2015epidemic}.
We quantify the fairness of the system in terms of how women acquire and diffuse knowledge compared to men.
For the acquisition of knowledge, we check whether the gender distribution of the authors reached by the spreading matches that of the whole collaboration network. 
For the diffusion of knowledge, we estimate whether women need more time than men to inform the same fraction of individuals in the network (\cref{fig:1}).

Our analysis of the collaboration network indicates that women in physics are disadvantaged in acquiring and diffusing knowledge.
Exploring the main determinants of this unfairness, we find that the gap in the number of women does not have a substantial impact on knowledge acquisition and diffusion; instead, the network structure intertwined with the dynamics of scientific collaborations contributes to relegating women to the margins.

\section*{Results}
\subsection*{Knowledge flow in time-varying collaboration networks}
\begin{figure}[t]
    \centering
    \includegraphics[width=.7\textwidth]{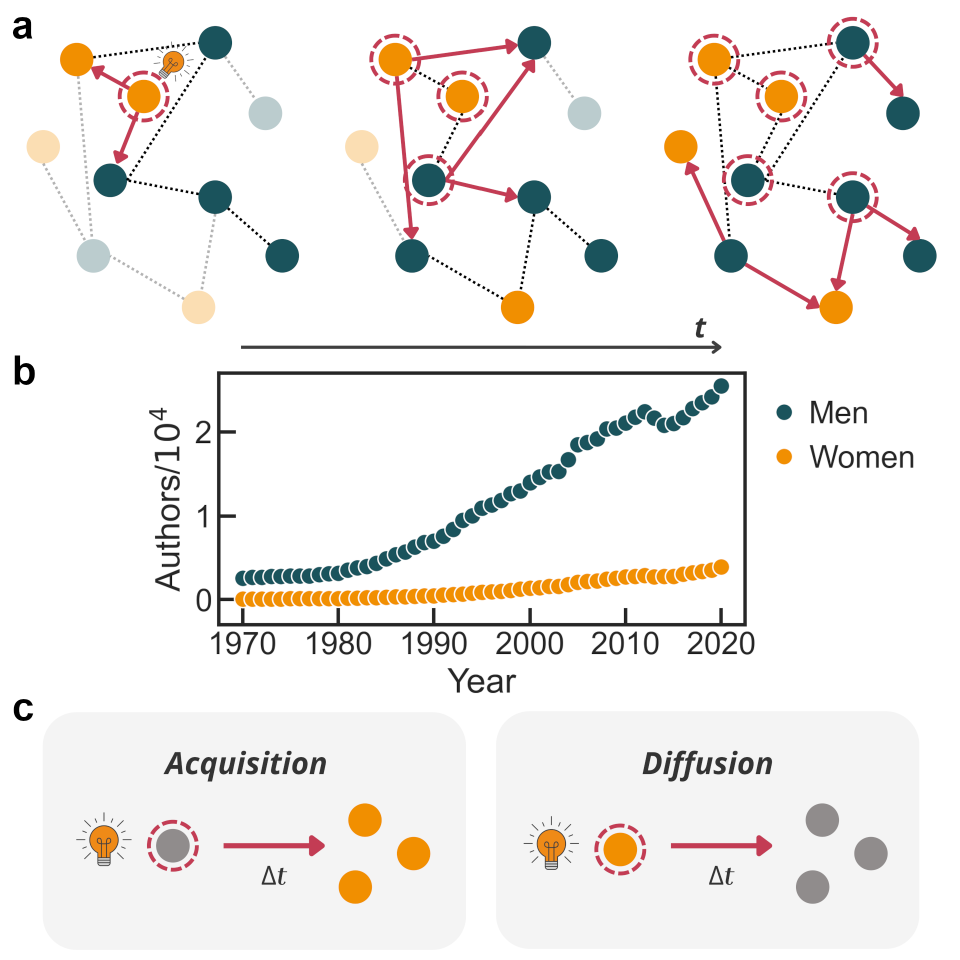}
    \caption{\textbf{Acquisition and diffusion fairness in the circulation of scientific knowledge.}
    \textbf{a} Schematic of knowledge flow through the collaboration network. 
    An author (node) starts sharing knowledge at a given year $t$, and the spreading continues following the author's connections (links).
    Once reached by the knowledge flow, informed nodes (circled in red) start diffusing to their uninformed neighbors.
    Authors are split into two groups according to their gender: Women (yellow) and men (blue).
    The network varies in time: At each year, new nodes enter the network and connections may change.
    \textbf{b} Number of authors over time. 
    Both male and female scholars have a growing trend.
    Women are the minority in the network, although their percentage has increased over the years.
    \textbf{c} Acquisition and diffusion of knowledge. 
    To quantify how women acquire and diffuse knowledge, we evaluate (i) the fraction of female authors reached when the spreading starts from a random (either male or female) author in the network (\textit{Acquisition}), and (ii) the fraction of authors reached when knowledge spreads from a woman (\textit{Diffusion}).
    }
    \label{fig:1}
\end{figure}

We analyze a large dataset of 678,916 scientific papers published in the journals of the American Physical Society (APS) between 1893 and 2020.
Each article comes with metadata on the date of publication, the papers cited, and authors' information, particularly full names and affiliations.
However, the APS dataset contains neither a unique identifier of the authors nor their gender.
Hence, we use a name disambiguation method to extract the identity of scientists and match them with their publications~\cite{sinatra2016impact,bachmann2024cumulativeadvantagebrokerageacademia}, and we infer gender based on author names~\cite{VanBuskirk_Clauset_Larremore_2023} (see Methods for details).
The final sample consists of 34,596 female and 195,490 male authors, plus 200,669 scientists for whom no reliable gender estimate was possible.

We build a coauthorship network, where nodes represent scientists and links represent collaborations among them, i.e., having coauthored a paper.
We leverage the temporal information of our dataset to build a time-varying collaboration network (\cref{fig:1}\textbf{a}).
Specifically, the network consists of a series of snapshots, each representing a publication year, where two authors are connected if they have coauthored a paper during that year.
The coauthorship network grows in time, as new nodes are added at every snapshot.
Specifically, when a scientist publishes for the first time in a given year, a node is added to the corresponding snapshot. 
Once in the network, nodes are not removed, hence isolated nodes in a certain snapshot represent authors that do not collaborate with others in that year.
For the following analyses, we only consider nodes that are part of the largest connected component of the network that aggregates all publication years from 1970 to 2020.
The final network consists of 367,836 nodes: 21,801 female and 135,552 male scholars, while the rest are scientists for whom gender is unknown.
We focus on this time interval because before the 70s the number of authors, and in particular the number of women, is too small for our analysis.

At a given year $t$, the network has $N_f(t)$ nodes representing female authors, $N_m(t)$ nodes representing male authors, and $N_u(t)$ nodes representing authors without gender information (unknown).
We denote as $N(t) = N_f(t) + N_m(t)$ the number of nodes at time $t$ for which gender is assigned.
Focusing on nodes with gender information, we observe that women are a minority in the network (\cref{fig:1}\textbf{b}), although their percentage has increased over the years (from 2.4\% in 1970 to 15.3\% in 2020).

We use the coauthorship network because we assume that scientific collaborations represent a form of social relationships through which (tacit) knowledge can flow \cite{singh2005collaborative,sorenson2006complexity,venturini2024collaboration}.
Furthermore, considering the temporal nature of the collaboration network 
provides a more realistic model for studying the flow of knowledge, as ties can change from one year to another, and those connections formed at one time and never repeated are reasonably less effective than persistent ones in conveying information \cite{breschi2009mobility}.

We model knowledge sharing as a simple contagion process on the time-varying network.
We assume that each node can be susceptible (S) or informed (I).
Susceptible nodes represent individuals that have not acquired the information flowing in the network, while informed nodes represent those who have.
At each time, informed nodes diffuse knowledge, making all of their susceptible neighbors informed.
Informed nodes do not become susceptible again, meaning that they do not stop spreading the knowledge they acquired (\cref{fig:1}\textbf{a}).
Although this model cannot capture the complexity of how knowledge is transmitted from one individual to another, its simplicity allows us to focus on how the structure and dynamics of the coauthorship network affect the flow of knowledge \cite{barrat2013temporal,vestergaard2014memory}.

\subsection*{Measuring acquisition and diffusion fairness}

To study gender disparities in knowledge dissemination, we select a publication year, $t_0$, choose one node in the corresponding network snapshot and make it informed, while leaving all other nodes susceptible.
All nodes that are added to the network in a given year $t > t_0$ will also be susceptible.
We run the contagion process following the temporal sequence of collaboration ties, and monitor at each year $t$ the number of informed women, $I_f(t)$, and informed men, $I_m(t)$, whose sum we denote as $I(t) = I_f(t) + I_m(t)$.

The system can be unfair towards female scholars in two directions (\cref{fig:1}\textbf{c}).
On the one hand, due to their number or position in the network, women may be unable to acquire information as quickly as men (Acquisition).
On the other hand, a contagion process starting from a woman could take more time to reach the same number of individuals than a process originating from a man (Diffusion).
\begin{figure}[t]
    \centering
    \includegraphics[width=.8\textwidth]{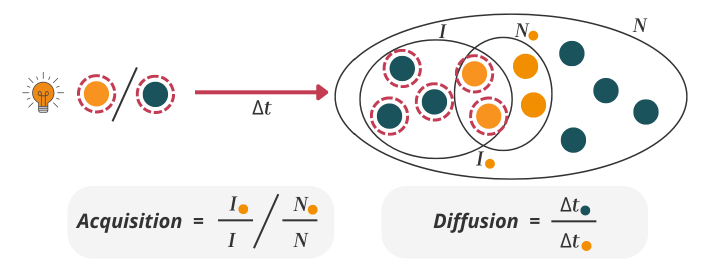}
    \caption{\textbf{Acquisition and diffusion fairness}.
    Knowledge can spread either from a female (yellow) or a male (blue) author.
    After a time $\Delta t$, among the $N$ authors of the collaboration network, $I$ of them will be informed (red circled).
    For the acquisition, we compare the fraction of informed individuals that are women to the fraction of female authors.
    For the diffusion, we compare the time $\Delta t$ needed to reach a given fraction of the population when the spreading starts from a woman or a man.
    }
    \label{fig:2}
\end{figure}

To evaluate fairness in knowledge acquisition, we let the spreading start from a randomly selected node (male or female, but not unknown) and let the process run until a given fraction of individuals $i_\mathrm{end}$ is reached in year $t_\mathrm{end}$.
Then, we count the fraction of informed nodes representing female authors, $i_f = I_f(t_\mathrm{end})/I(t_\mathrm{end})$, and the fraction of women in the network, $n_f = N_f(t_\mathrm{end})/N(t_\mathrm{end})$ (\cref{fig:2}).
Following the principle of statistical parity \cite{mehrabi2021survey}, we define \textit{acquisition fairness} as
\begin{equation}
    AF = \left\langle\frac{i_f}{n_f}\right\rangle,
\end{equation}
where $\langle\cdot\rangle$ indicates an average over multiple repetitions of the contagion process, i.e., multiple draws of the initial spreader, removing the dependence on the particular node selected.
When $AF < 1$, the fraction of informed individuals who are female is smaller than the fraction of women, meaning that there are fewer informed women than one would expect, i.e., the system is unfair towards the minority.
Conversely, $AF > 1$ means that less men are exposed to knowledge flow compared to what we would expect, while with $AF \approx 1$ the system is fair in knowledge acquisition.
Note that, for a given starting year $t_0$, the acquisition fairness varies as a function of the target fraction of informed individuals $i_\mathrm{end}$.

To investigate fairness in knowledge diffusion, we compare the spreading processes starting from women with those generated by men.
Again, we let both types of processes propagate until a certain fraction $i_\mathrm{end}$ is reached in year $t_\mathrm{end}$ and compute the length of each process, namely $\Delta t = t_\mathrm{end} - t_0$.
We define \textit{diffusion fairness} as 
\begin{equation}
    DF = \frac{\langle\Delta t_m\rangle}{\langle\Delta t_f\rangle},
\end{equation}
where the subscripts indicate the gender of the initial spreader.
$DF < 1$ indicates that knowledge flows initiated by women take on average more time to reach the same number of individuals than processes started by men, so the system is unfair to the minority group.
$DF > 1$ denotes, instead, an unfairness towards the majority, while a value $DF \approx 1$ suggests fairness in knowledge diffusion.

\subsection*{Women are disadvantaged in acquiring and diffusing knowledge}
\begin{figure*}[ht!]
\centering
\includegraphics[width=\textwidth]{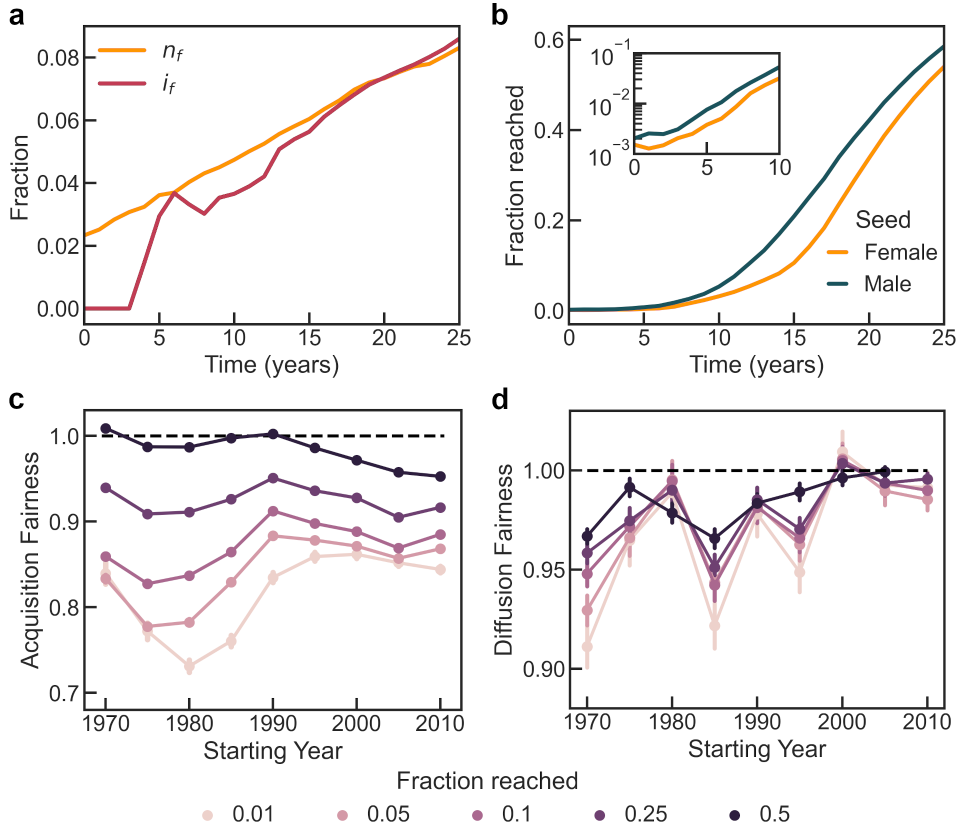}
\caption{
\textbf{Fairness in the acquisition and diffusion of knowledge.}
\textbf{a} Example of women's acquisition of knowledge for a contagion process that starts from a man.
The fraction of female scholars in the collaboration network, $n_f$ (yellow) is compared to the fraction of informed scientists that are women, $i_f$ (red).
Women among the informed individuals are generally less than expected.
\textbf{b} Example of how knowledge diffuses from a female author, compared to a process that starts from a male author.
The process starting from a woman takes longer than that starting from a man to reach the same fraction of individuals.
\textbf{c} Acquisition fairness as a function of the starting year of the knowledge flow.
The curves indicate different values of the fraction of individuals reached.
The dashed line represents the fair condition in which the fraction of women among those who acquired knowledge is equal to the fraction of women in the network.
In general, the acquisition fairness is below the fair condition, meaning that the women reached by the spreading are less than expected.
\textbf{d} Diffusion fairness as a function of the starting year for different values of the fraction of individuals reached.
The dashed line represents the fair condition in which the average duration of the process is the same when starting either from a woman or a man.
In general, the diffusion fairness is close but significantly below the fair condition, meaning that women take slightly longer than men to diffuse their knowledge.
}
\label{fig:3}
\end{figure*}
We first show an example of how the system can be unfair to women in knowledge acquisition and diffusion.
\cref{fig:3}\textbf{a} shows the evolution of the fraction of women in the collaboration network, $n_f$ (yellow), and the fraction of informed individuals who are women, $i_f$ (red), for a process starting in 1970 from a male individual.
We observe that $i_f=0$ for the first years of the spreading, meaning that all researchers reached by the knowledge flow in that time range are men.
After that initial phase, information also reaches women, and $i_f$ starts to rise.
After more than 15 years, $i_f$ reaches $n_f$.
That is, the fraction of informed researchers who are women is the same as the fraction of women in the collaboration network.
\cref{fig:3}\textbf{b} shows the evolution of the fraction of informed nodes (both male and female), for two different processes starting in 1970 from a man (blue) and from a woman (yellow).
The process starting from the female node is slower than that starting from the male node, with a gap ranging from one to four years for a given fraction of informed individuals. 
While this difference widens in time, we observe that it already emerges at the early stages of the spreading (inset of the panel, focusing on the first 10 years of the process).
This suggests that women are disadvantaged in diffusing their knowledge, as they need more time than men to reach the same number of researchers in the network.

We now perform a systematic analysis of the system unfairness using the measures we have introduced, beginning with an evaluation of the acquisition fairness.
We consider a publication year $t_0$, and let the knowledge spread until a target fraction of informed individuals $i_\mathrm{end}$ is reached. 
We perform 1000 simulations, each time randomly selecting the node from which the spreading starts.
The acquisition fairness is evaluated as the average over those repetitions. 
We consider different starting years, from 1970 to 2010 at steps of five years, ignoring starting years after 2010, so that the process can last for at least ten years.
We estimate the acquisition fairness for different values of $i_\mathrm{end}$, ranging from $1\%$ to $50\%$ of the nodes for which we have gender information.
In this way, we assess the fairness of the system among those who acquire knowledge at the early stages of the diffusion process (e.g., first 1\%) and those reached later on.

Results are shown in \cref{fig:3}\textbf{c}.
Each curve represents the acquisition fairness as a function of the starting year $t_0$, for different values of the fraction of individuals to be reached $i_\mathrm{end}$.
The dashed line at $AF = 1$ shows the fair condition, i.e., the final fraction of women reached is equal to the fraction of women present in the network at that time.
In general, the acquisition fairness is considerably below the fair condition, which means that women reached by the knowledge flow are less than expected.
This is especially pronounced for those who acquire knowledge at the early stages of the contagion: For instance, when a process starts in 1980, women are on average 25\% less than expected among the first 1\% of informed individuals.
Such result can have far-reaching implications, as acquiring knowledge early in time can generate competitive advantages that eventually exacerbate gender disparities.
Besides early acquisition, the larger the fraction of individuals reached, the closer we get to the fair condition $AF = 1$.
In addition, we note that the acquisition fairness has a non-trivial dependence on the starting year.
While we observe a U-shaped trend between 1970 and 1990, we note a decreasing trend afterwards, which supports the previous literature questioning whether the simple increase in the number of women is sufficient to reduce gender differences in physics \cite{neuhauser2023improving}.

Focusing on the diffusion fairness of the system, we run two different sets of simulations.
In the first, we fix a starting year $t_0$, choose a random female author, and let the process run until a given fraction of scientists $i_\mathrm{end}$ is reached.
We run 4000 simulations, each time picking a different node to initiate the diffusion.
In the second, we follow the same procedure, selecting a starting node that instead represents a male author.
Again, we repeat this 4000 times, choosing different nodes at the beginning of the contagion process.
We then evaluate the diffusion fairness $DF$, comparing the average duration of the contagion starting from a woman, $\Delta t_f$, to that starting from a man $\Delta t_m$.
We consider again different starting years ranging from 1970 to 2010 in five-year intervals, evaluating $DF$ across various $i_\mathrm{end}$, from $1\%$ to $50\%$.

\cref{fig:3}\textbf{d} shows the results of diffusion fairness.
Each curve portrays the diffusion fairness as a function of the starting year $t_0$, for different fractions of individuals to be reached $i_\mathrm{end}$.
The dashed line at $DF = 1$ shows the fair condition, in which the average duration of the diffusion process is the same when starting from a woman or a man.
Though the curves lie close to $DF = 1$, they differ significantly from it for most starting years, meaning that men can have an advantage in knowledge diffusion compared to women.
Moreover, we note that the curves tend to overlap, suggesting that the diffusion fairness does not vary appreciably as a function of $i_\mathrm{end}$.
In other words, regardless of the number of individuals that one aims to reach, women are slower than men in the dissemination of their knowledge through the collaboration network.

\cref{fig:3} gives an overall picture of women's disadvantages in acquiring and diffusing knowledge.
Specifically, among those who acquire knowledge early in time, women are less represented than men.
Moreover, women take slightly more time than men to diffuse their knowledge.
In the Supplementary Information we complement this analysis by investigating the contagion process on a time-aggregated version \cite{holme2012temporal} of the collaboration network (Fig.~S1).

\subsection*{The determinants of unfairness}
\begin{figure*}[ht!]
    \centering
    \includegraphics[width=\textwidth]{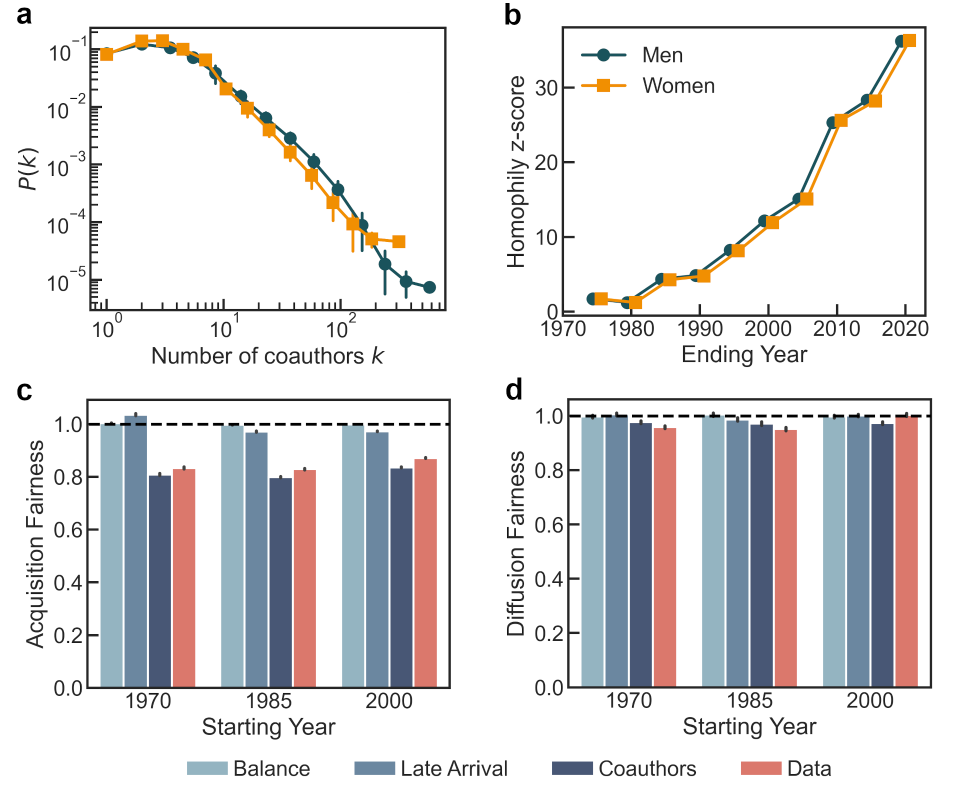}
    \caption{\textbf{Determinants of the unfairness of the system}. 
    \textbf{a} Probability distributions of the number of coauthors for men (blue) and women (yellow).
    On average, female scholars have less collaborators than male scholars.
    \textbf{b} Evolution of homophily in the APS collaboration network.
    The tendency of both male (blue) and female (yellow) scholars to collaborate with scientists having their same gender has increased over the decades.
    See Supplementary Information for details on how to evaluate homophily.
    \textbf{c}-\textbf{d} The determinants of acquisition and diffusion unfairness.
    Bars represent the value of acquisition and diffusion unfairness for the null models (shades of blue) and the data (red), for three starting years of the spreading process. 
    The ``Balance'' model shows no unfairness, i.e., $AF=1$ and $DF=1$ (black dashed lines).
    The ``Late Arrival'' model does not show unfairness, suggesting the gap in the number of women in the collaboration network does not impact the way women acquire and diffuse knowledge.
    The ``Coauthors'' is in good agreement with the data, hinting that the gap in the number of coauthors has a significant negative effect on acquisition and diffusion fairness.
    }
    \label{fig:4}
\end{figure*}
The unfairness in the acquisition and diffusion of knowledge could be determined by different factors.
First, there is an imbalance in the number of men and women in the system, which is not constant over time; indeed, women arrive later in the network, and this might affect their opportunities to receive and send information.
Additionally, women and men differ in their collaboration patterns (\cref{fig:4}\textbf{a}), as women have usually fewer coauthors over their careers \cite{zeng2016differences}.
This gap could contribute to the acquisition and diffusion unfairness.
We focus on the influence of those factors by designing three null models of the collaboration network where we assign gender to individuals so as to preserve (or remove) certain characteristics of the system (see Methods for details).
The first null model (to which we refer as ``Balance'') generates a scenario where women are 50\% of the researchers at any point in time and women and men have the same collaboration patterns, i.e., the same degree distributions in the time-aggregated network \cite{holme2012temporal}.
With the second null model (``Late Arrival'') we preserve how the fraction of women in the network evolves while removing gender differences in the number of collaborators, which means, again, that women and men have the same time-aggregated degree distributions.
Finally, the third model (``Coauthors'') preserves the proportion of women over time and the difference in the number of coauthors between men and women, potentially removing other gender-specific patterns in the data, e.g., homophily \cite{jadidi2018gender}.

For each model, we consider 1970, 1985, and 2000 as starting years, and let knowledge flow until a target fraction of informed individuals $i_\mathrm{end}$ is reached. 
We consider $i_\mathrm{end}=0.05$ for acquisition fairness, while $i_\mathrm{end}=0.25$ for diffusion fairness.
This choice remarks on the different potential advantages in knowledge dissemination: Researchers benefit from acquiring knowledge earlier in time and from diffusing it to as many peers as possible.
We evaluate the acquisition fairness as the average over 1000 simulations (10 network randomizations, 100 runs of the spreading per randomization), in which the starting node is selected at random each time.
For the diffusion fairness, we compare the average spreading time of 5000 simulations starting from a random woman to the average spreading time of 5000 simulations starting from a random man (in both cases, 50 network randomizations, 100 runs of the spreading per randomization).

\cref{fig:4}\textbf{c}-\textbf{d} show the results obtained for acquisition and diffusion, respectively.
In the ``Balance'' model, there is no acquisition or diffusion unfairness, i.e., $AF=1$ and $DF=1$, no matter the starting year.
As the model generates a scenario in which there are no appreciable differences between men and women, this result does not come as a surprise and acts as a validation of our measures of fairness. 

Remarkably, also the ``Late Arrival'' model does not show unfairness. 
This suggests that neither the gap in the number of women nor its temporal evolution has an impact on the way female researchers acquire and diffuse knowledge.
In the case of acquisition, the flow of knowledge reaches more men than women ($i_f < i_m)$ but almost in the same proportion as the number of men and women in the network ($i_f/n_f \approx i_m/n_m$); therefore, the system is fair.
For diffusion, instead, the number of women does not play a significant role, as we compare the average time that men and women need to inform a certain fraction of individuals, regardless of their gender.

Unfairness emerges once we account for network features.
Specifically, the ``Coauthors'' model proves that the gap in the number of collaborators has a significant negative effect on the fairness of the system.
The difference in coauthor distributions seems to explain most of the unfairness, as the model is in good agreement with empirical data.
The analysis of this model shows how the structure of the collaboration network weakens the ability of women to acquire and diffuse knowledge from and to their peers.

The small mismatch we observe between this last model and the data suggests that other network features might affect the system fairness.
First, the distribution of the number of collaborators for both genders is not constant over time (see Fig.~S2 in Supplementary Information).
The ``Coauthors'' model flattens out these temporal differences, since it only controls for the distributions in the time-aggregated version of the network.
Second, beyond the number of coauthors, the exact position of the nodes in the network can add to the system unfairness (see Fig.~S3 in Supplementary Information).
Finally, our null model shuffles the gender of the nodes without controlling for homophily. 
However, gender homophily in scientific collaborations substantially increased for both men and women over the years (\cref{fig:4}\textbf{b}; see Supplementary Information for details on how we measure homophily).
Therefore, it is likely that homophily generates a gap between the null model and the data.

Together, the analysis of the null models sheds light on the factors underlying gender differences in the acquisition and diffusion of knowledge over time.
Specifically, we find that unfairness is not driven by the gap in the number of women, rather it arises from the structure of the collaboration network that creates a cumulative disadvantage for women over time.

\section*{Discussion}
In this work, we investigated gender disparities in the dissemination of scientific knowledge through the APS collaboration network.
Although, in principle, knowledge flow is not contingent on gender, we discovered a systematic trend suggesting that women may face challenges in acquiring and diffusing knowledge.
These disparities are critical: On the one hand, late knowledge acquisition can generate a competitive disadvantage that may prevent women from accessing novel ideas, new collaborations, or even finding employment; on the other hand, slower knowledge diffusion may negatively affect women's visibility, fostering gender stereotypes and impacting their academic recognition, exacerbating the Matilda effect \cite{rossiter1993matthew}.
Our findings suggest that these two dimensions of unfairness are not symmetric: Whereas the difference in knowledge diffusion is less marked and seems to improve over time, women's disadvantage in knowledge acquisition is considerable and persistent.

Exploring the potential drivers of unfairness, we found that the gap in the number of female scholars does not explain women's disadvantage; instead, the difference in the cumulative number of collaborators jeopardizes women's chances to acquire and diffuse knowledge.
Beyond knowledge dissemination, different collaboration patterns could limit opportunities for women throughout their careers, as job and promotion offers are often the result of peer-to-peer interactions.

These findings challenge the hypothesis that increasing the number of women in STEM is enough to provide equal opportunities to acquire and diffuse scientific knowledge.
This lack of opportunities is not simply related to \textit{when} women enter the system but to \textit{how} they take part in it, that is, the evolution of their network of collaborators.
Differences between women and men in the size of their networks may stem from multiple factors.
First, as a historically disadvantaged minority in a male-dominated field and with more parental responsibilities, women might be excluded from cutting-edge collaboration opportunities \cite{sugimoto2023equity,morgan2021unequal,cech2019changing}. 
Second, women are more likely to drop out of academia in all career ages compared to men \cite{huang2020historical}, preventing them from growing their network (see Fig.~S4 in Supplementary Information for the analysis of career lengths in our data).
Moreover, women tend to have less academic mobility than men \cite{momeni2022many}: Since mobility plays a crucial role in expanding scholars' network, this might contribute to widening the gap in the number of collaborators. 
Remarkably, these differences are intertwined with time, which results in slowing down the acquisition and diffusion of knowledge, thus creating
a temporal glass ceiling \cite{cotter2001glass} for women in physics.

Our work sheds light on how the dynamics of scientific collaborations shape gender disparities in knowledge dissemination and calls for a deeper understanding on how to intervene to guarantee fairness in the scientific community.
Specifically, it highlights that policies intended to increase the number of female scholars in academia may not be sufficient to reduce gender disparities in the acquisition and diffusion of scientific and technical knowledge.
In line with the policy changes recently suggested by the scientific community \cite{greider2019increasing,diele2021potential,beidas2022advancing,teichmann2022community}, our findings point to institutional interventions that prevent women's career dropout and encourage academic mobility and diverse collaborations, with the aim of promoting equity, diversity, and inclusion in science.

\section*{Materials and Methods}
\subsection*{Name disambiguation and gender inference in APS dataset}
We construct the collaboration network from 678,916 publication records of the American Physical Society, ranging from $1893$ to $2020$.
This dataset contains, for each publication, a list of coauthor names and their affiliations, alongside other metadata.
We apply a recently developed implementation~\cite{bachmann2024cumulativeadvantagebrokerageacademia} of an existing name disambiguation solution~\cite{sinatra2016impact} to link publications to authors.
For a pair of publications with matching author names, this solution uses affiliations, citations, and coauthors to decide whether the two names and the respective publications belong to the same author.
It applies a conservative strategy by merging only authors for which there are no conflicting first names.
This reduces the number of authors with extreme publication counts.
In the collaboration network, this limits the number of nodes with very high degree, which could deeply affect the contagion dynamics.

Similar to previous studies~\cite{kong2022influence,bachmann2024cumulativeadvantagebrokerageacademia,huang2020historical}, we infer gender labels from author names.
Using a recently developed open-source solution~\cite{VanBuskirk_Clauset_Larremore_2023}, 
we estimate 34,596 scientists to be women, 195,490 to be men, while the gender of 200,669 disambiguated authors remains unknown.
For authors with multiple names, we remove names for which no gender could be inferred and assign gender based on a majority vote, i.e., if the majority of names are labeled as male (female), the scientist is considered male (female).
Authors with a draw in the number of female and male labeled names or those for whom only ``unknown'' labels could be inferred, most often due to providing only initials or unisex names, are labeled as ``unknown''.
Note that these authors remain present in the network, namely they participate in the knowledge spreading process.
However, they are ignored when 1) seeding the contagion process and 2) evaluating the fairness of the system.

\subsection*{Null models of time-varying collaboration networks}
We explore the factors underlying the emergence of acquisition and diffusion unfairness in the flow of knowledge by designing three null models of the APS collaboration network.
For each model, the structure of the temporal network, namely the nodes and the connections between them in every snapshot, is unchanged.
In contrast, the node attributes, meaning the authors' gender, are reshuffled to preserve certain characteristics of the system.

\begin{itemize}
    \item \textbf{Balance}. This model generates an ideal scenario where 1) 50\% of the authors are women at any time; 2) men and women do not differ in any network feature.
    Starting from the first snapshot of the temporal network, we randomly assign the gender to the nodes with a 50/50 ratio.
    In the following snapshots, old nodes keep their gender, while the gender of new nodes is again assigned at random with a 50/50 ratio.
    The main purpose of this model is to validate our measures of fairness, as the scenario it creates should be perfectly fair.
    \item \textbf{Late arrival}. This model preserves the proportion of women in each year so that it remains as it is in the data and removes gender differences in the network features.
    We reshuffle the gender of the nodes in the first snapshot of the network.
    In the following snapshots, we keep the gender of old nodes and reshuffle that of new ones.
    The model allows us to investigate how the late arrival of women in the collaboration network affects their opportunity to acquire and diffuse scientific knowledge, while controlling for the structure of the network.
    \item \textbf{Coauthors}. With this model, we preserve the proportion of women in each year and the difference between genders in the distribution of the number of coauthors.
    First, we order nodes by their total number of connections and divide them into groups, each representing 2\% of the nodes.
    Analogously to the Late arrival model, we start from the first snapshot and reshuffle the gender of nodes within the same group.
    In this way, we explore how gender differences in the distribution of coauthors impact the fairness of the system on top of the late arrival of women in the network.
\end{itemize}

\bibliography{biblio}

\bibliographystyle{ScienceAdvances}

\section*{Acknowledgments}
C.Z. acknowledges funding by the European Union under Horizon EU project LearnData, 101086712.
C.Z. and L.G. thank C\'esar Hidalgo, Bal\'azs Lengyel, Vito Latora and Esteban Mu\~noz for the insightful comments and discussion.
L.G. and F.B. acknowledge support of the Air Force Office of Scientific Research under award number FA8655-22-1-7025.
J.B. was supported by the Austrian Science Promotion Agency FFG under project No. 873927 ESSENCSE and the DOC Fellowship of the Austrian Academy of Sciences.

\section*{Competing interests} The authors declare they have no competing interests.

\section*{Data and material availability}
The APS datasets are available upon request to the American Physical Society \\(https://journals.aps.org/datasets).

\end{document}



\maketitle


\section*{Analysis of the static network}
In the main text, we investigated the unfairness in knowledge dissemination on a collaboration network that evolves in time. 
Here we complement this analysis by studying a static, time-aggregated version of the collaboration network~\textit{(43)}. 
Specifically, two authors in this network will be connected if they coauthored at least one paper together between 1970 and 2020.
This network cancels out all temporal features of the system: The fraction of women does not vary (female scholars are 16.1\%), the network does not grow, and collaborations have neither dynamics (i.e., links are always active) nor temporal order (i.e., links represent collaborations occurring in any year).

Again, we model knowledge sharing as a diffusion process where nodes can be either susceptible (S) or informed (I).
At each time, informed nodes can make their susceptible neighbors informed.
While for the temporal network we assumed that infections occur with probability $\beta_T = 1$, we will consider a probability $\beta_S < 1$ for the static network.
The reason is that the average degree in each snapshot of the temporal network is smaller than the average degree in the static network, i.e., $\langle k_T\rangle(t) < \langle{k_S}\rangle$.
Therefore, with $\beta_S = 1$, the spreading in the static network will be substantially faster than the spreading in the time-varying network.
To allow for a fair comparison between the two networks, we consider a value of $\beta_S$ for which the average number of newly informed individuals (at an early stage of the spreading) is the same for the two networks.
In the static network, this is given by $\beta_S \langle{k_S}\rangle I$, while in the temporal network this is given by $\langle{k_T}\rangle I$, where $\langle{k_T}\rangle$ is the average value of $\langle k_T\rangle(t)$ over the network snapshots, and $I$ is the number of informed individuals.
Assuming these two quantities to be equal, we finally obtain 
\begin{equation}
    \beta_S = \frac{\langle{k_T}\rangle}{\langle{k_S}\rangle} = \frac{1}{49\langle{k_S}\rangle}\sum\limits_{t=1970}^{2020}\langle k_T \rangle(t)
\end{equation}

We first evaluate the acquisition fairness. 
As in the main text, we select a starting node and let the knowledge spread until a fraction of informed individuals $i_\mathrm{end}$ is reached. 
To fairly compare the time-varying and time-aggregated versions of the network, we consider authors that have published in 1970, and let the diffusion process start from the nodes corresponding to those authors.
We perform 1000 simulations, each time randomly selecting the node from which the spreading starts.
The acquisition fairness is then evaluated as an average over these repetitions. 
We evaluate how the acquisition fairness as a function $i_\mathrm{end}$, ranging from $1\%$ to $50\%$ of the nodes for which we have gender information.

\begin{figure*}[t!]
    \centering
    \includegraphics[width=\textwidth]{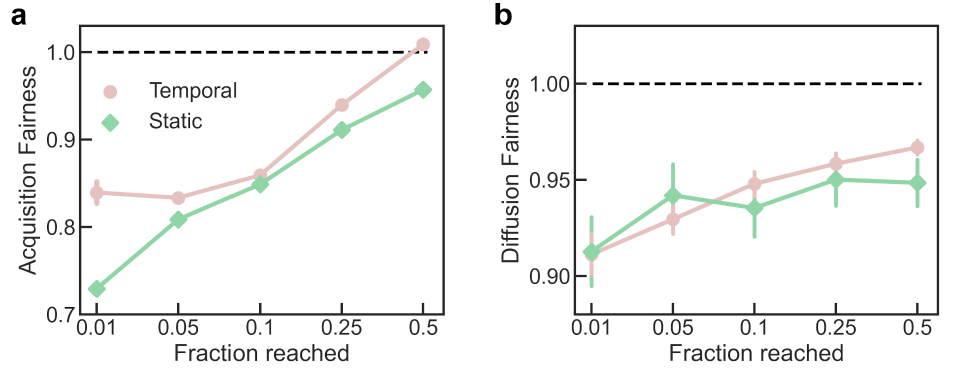}
    \caption{\textbf{Acquisition and diffusion fairness in the time-varying and in the time-aggregated network}. Panel \textbf{a} shows the acquisition fairness evaluated for the temporal (pink) and the static (green) versions of the collaboration network.
    Panel \textbf{b} shows analogous results for the diffusion fairness.}
    \label{fig:s1}
\end{figure*}

\cref{fig:s1}\textbf{a} shows the comparison between the time-varying (pink) and the time-aggregated (green) versions of the collaboration network.
For the static network the acquisition fairness is significantly less than 1 (i.e., fair condition), proving that women are disadvantaged compared to men in acquiring scientific knowledge regardless of the temporal nature of the collaboration network.
In particular, the value of $AF$ for the static network is always smaller than the value obtained on the temporal network.
This is likely due to how we define the acquisition fairness (see Results in the main text).
In particular, this compares the fraction of informed individuals that are women to the fraction of women in the network.
In temporal network the latter varies in time, while in the static network it is a constant. 
Furthermore, the fraction of women in the time-aggregated network is larger than the fraction of women in every snapshot of the time-varying network.
Therefore, the expected fraction of women within the informed authors is always larger and thus the acquisition fairness is systematically lower.

We now focus on the diffusion fairness of the system.
We run two different sets of simulations.
In the first, we choose a random node corresponding to a female author and let the process run until a fraction of individuals $i_\mathrm{end}$ is reached.
We run 1000 simulations, each time picking a different node to initiate the diffusion.
In the second, we follow the same procedure, selecting a starting node that instead corresponds to a male author.
Again, we repeat this a 1000 times, choosing different starting nodes.
We evaluate the diffusion fairness comparing the average duration of the process starting from a woman $\Delta t_f$ to that starting from a man $\Delta t_m$.
As for the acquisition fairness, we only select authors present in 1970 and analyze diffusion fairness as a function of $i_\mathrm{end}$, from $1\%$ to $50\%$.

The comparison between the time-varying (pink) and time-aggregated (green) networks is displayed in \cref{fig:s1}\textbf{b}.
The diffusion fairness is less than 1 (i.e., fair condition) for every fraction reached, hallmarking that women are disadvantaged compared to men notwithstanding the dynamics of the collaboration network.
The value of $DF$ for the two versions of the network is not significantly different, suggesting that the temporal correlations in the network structure are not the main determinants of the emergence of diffusion unfairness in the system.

\section*{Temporal evolution of the coauthor distributions}
In this section, we study how the number of collaborators change over time for male and female scientists.
Specifically, we consider a publication year $y_{end}$ and build a network where nodes represent authors that have published at least a paper between 1970 and $y_{end}$, and links connect authors that have collaborated at least once within the same time range.
Then, we evaluate the degree distributions, i.e., the distributions of the number of collaborators, for men and women separately. 
We consider different values of $y_{end}$, namely 1980, 1990, 2000, 2010, and 2020, so as to explore the temporal evolution of the coauthor distributions.
The year 2020 recovers the distributions showed in Fig.~4\textbf{a} of the main text.

\begin{figure*}[t!]
    \centering
    \includegraphics[width=.98\textwidth]{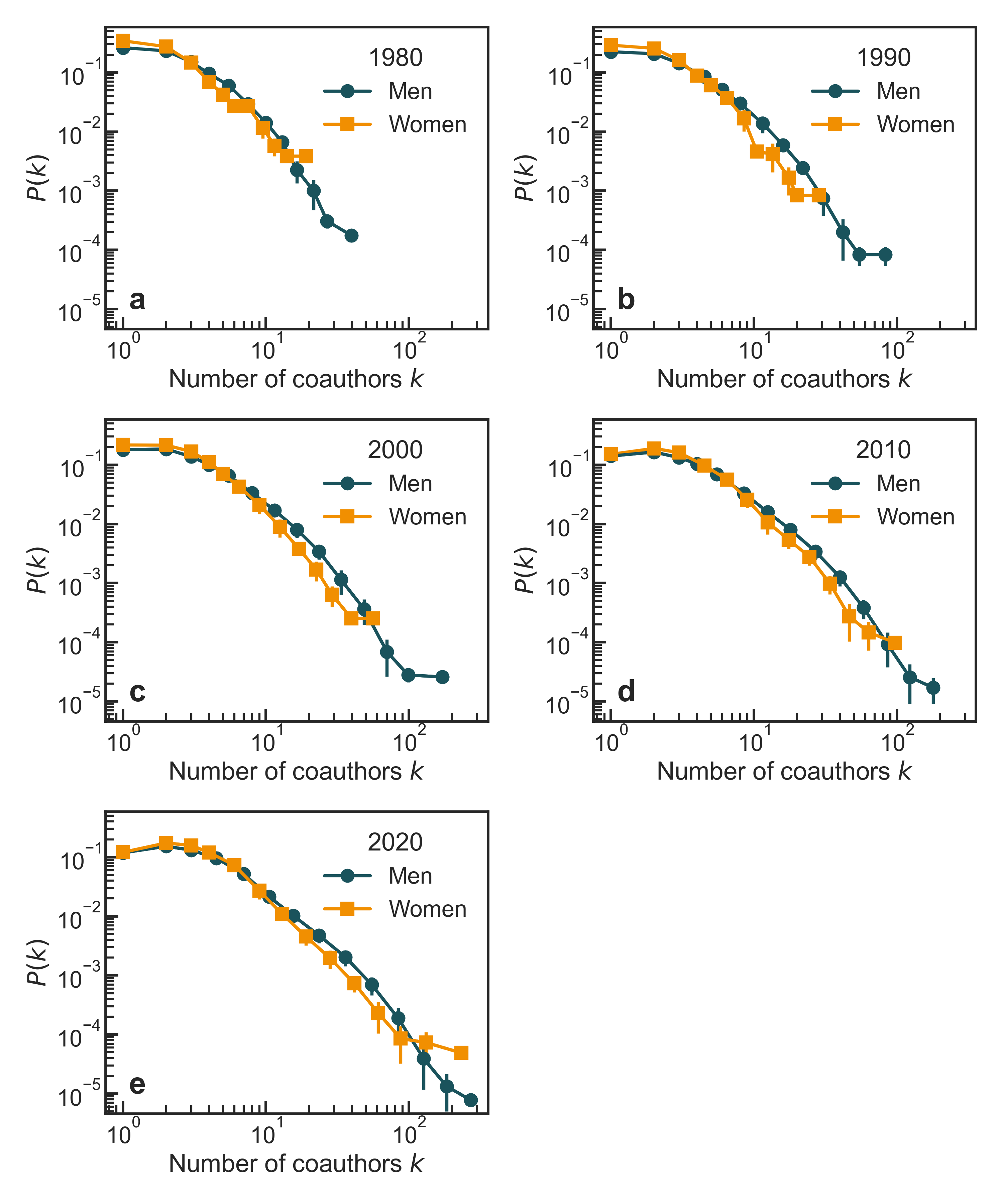}
    \caption{\textbf{Evolution of the coauthor distributions for men and women.} Panels \textbf{a}-\textbf{e} show the coauthor distributions for men (blue) and women (yellow) considering all articles published from 1970 to different ending years (one for each panel), reported on top of panel legends.}
    \label{fig:s2}
\end{figure*}

\cref{fig:s2} depicts how the distributions of collaborators for men and women evolve over time. 
In general, the distributions broaden in time regardless of gender, meaning that both male and female scholars had increased the size of their collaboration networks.
Crucially, the two distributions differ for every ending year. 
In particular, we note that women have in general narrower distributions, meaning that researchers with the largest number of collaborators are men.
Most importantly, women have overall fewer collaborators than men.
We quantify the difference between the two distributions using the Mann-Whitney U test, confirming that the distribution of coauthors for men is significantly greater than that for women for every ending year (p-values are reported in \cref{tab:s1}).

\begin{table}[h]
    \centering
    \begin{tabular}{c|c}
         $y_{end}$ & p-value \\
         \hline
         1980 & $1.8\cdot 10^{-5}$ \\
         1990 & $1.9\cdot 10^{-22}$ \\
         2000 & $9.1\cdot 10^{-42}$ \\
         2010 & $1.3\cdot 10^{-54}$ \\
         2020 & $3.5\cdot 10^{-89}$ 
    \end{tabular}
    \caption{\textbf{p-values of the Mann-Whitney U test on the coauthor distributions}. The low p-values indicates that the distribution of coauthors for men is significantly greater than that for women in all possible time range from 1970 to $y_{end}$.}
    \label{tab:s1}
\end{table}

\clearpage

\section*{The impact of closeness on acquisition and diffusion fairness}

\begin{figure*}[p!]
    \centering
    \includegraphics[width=\textwidth]{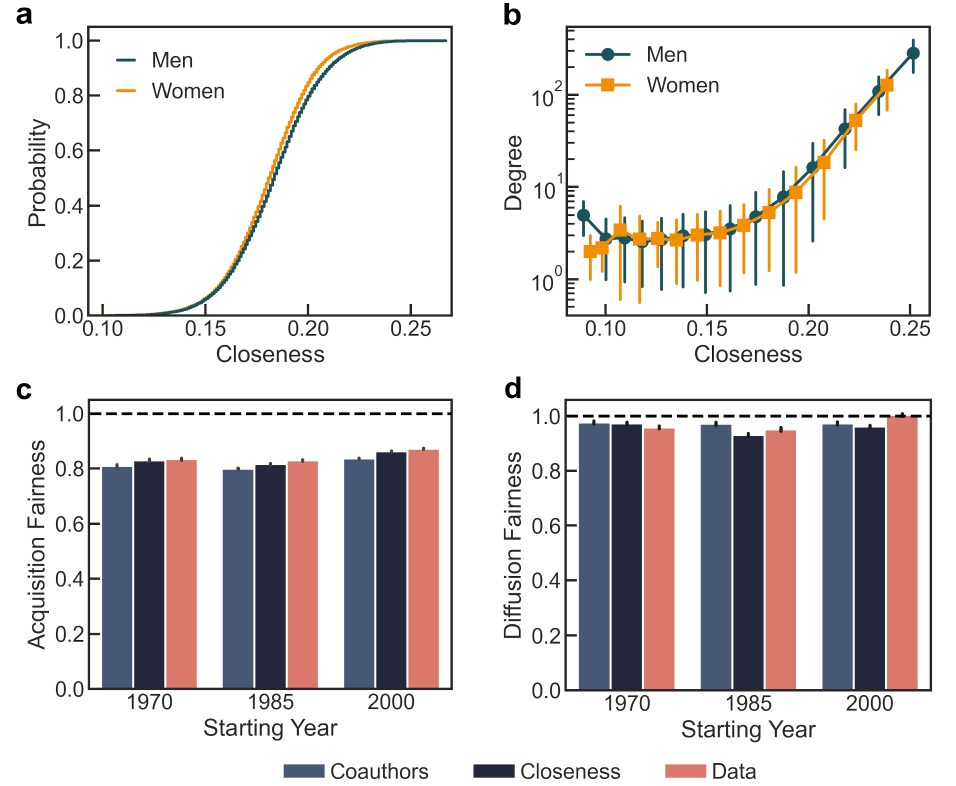}
    \caption{\textbf{The impact of closeness centrality on acquisition and diffusion fairness}. Panel \textbf{a} shows the cumulative distribution of the closeness centrality, for men (blue) and women (yellow) respectively. Panel \textbf{b} highlights the relationship between closeness and degree (i.e., number of coauthors) in the time-aggregated network, again for both men (blue circles) and women (yellow squares). Panels \textbf{c-d} present the results for the acquisition and diffusion, showing that both the ``Coauthors'' (dark blue) and ``Closeness'' (black) models are in agreement with the data (orange).}
    \label{fig:s3}
\end{figure*}

In the main text, we investigated what factors can limit women's opportunities of acquiring and diffusing scientific knowledge.
Here, we study how gender differences in closeness centrality impacts the fairness of the system.
Closeness, which measures the mean distance from a node to other nodes \cite{newman2018networks}, is not evenly distributed in the collaboration network, as men are generally closer than women to other authors (cumulative distributions for men and women are shown in \cref{fig:s3}\textbf{a}; we verify that the distributions are significantly different using a Mann-Whitney U test ($p=6\cdot 10^{-59}$). 
Being less central in the collaboration network, women can thus be disadvantaged in acquiring and diffusing knowledge. 

To study the impact of closeness, we design a null model (``Closeness'') that randomly assigns gender to the nodes by preserving 1) the temporal evolution of the fraction of female scholars in the network and 2) the closeness centrality of men and women in the static network.
First, we group nodes into percentiles based on their closeness.
We start assigning genders from the first snapshot of the temporal network, reshuffle the genders of nodes within the same percentile.
In the following snapshots, we keep the gender of old nodes and reshuffle that of new ones.

We consider three publication years, i.e., 1970, 1985, and 2000, and let the knowledge flow until the target fraction of informed individuals $i_\mathrm{end}$ is reached, considering $i_\mathrm{end}=0.05$ for acquisition and $i_\mathrm{end}=0.25$ for diffusion.
The acquisition fairness is evaluated as the average over 1000 simulations (10 network randomizations and 100 runs of the spreading process per each randomization), selecting a random starting node each time.
The diffusion fairness, instead, is calculated by comparing the average spreading time of 5000 simulations starting from a random woman to the average spreading time of 5000 simulations starting from a random man  (50 network randomizations and 100 runs of the spreading process per each randomization).

\cref{fig:s3}\textbf{c}-\textbf{d} display the results for acquisition and diffusion fairness, respectively.
In both panels, we compare the ``Coauthors'' (preserving the degree in the static network, see Methods) and ``Closeness'' null models to the data.
The ``Closeness'' model hints that the gender difference in the centrality has a negative effect on the fairness of the system.
Moreover, the ``Coauthors'' and the ``Closeness'' models are both in good agreement with empirical data.
This may be due to the fact that node degree and closeness are usually highly correlated \cite{evans2022linking}, as we have also verified for the APS collaboration network (see \cref{fig:s3}\textbf{b}).

The analysis of the closeness further highlights how the position of female scholars in the collaboration network can limit their opportunities to acquire and diffuse scientific knowledge.

\clearpage

\section*{Measuring homophily}
In this section we detail how to evaluate the evolution of homophily in the APS collaboration network.
Following~\textit{(39)}, for each snapshot of the temporal network we consider a null model that preserves the network size and the degree of the nodes, while randomizing node connections.
As networks generated with this configuration model \cite{newman2018networks} are not characterized by homophilic preferences, we can study the homophily of each snapshot by comparing it to these synthetic networks.
Specifically, for each publication year, we generate 100 networks and evaluate the average fraction of links connecting two women ($e_{ff}$), or two men ($e_{mm}$), and the corresponding standard deviations ($\sigma_{ff}$ and $\sigma_{mm}$).
Given the observed fraction of links $h_{ff}$, and $h_{mm}$, we assess a set of z-scores
\begin{equation}
    z_{xy} = \frac{h_{xy}-e_{xy}}{\sigma_{xy}}
\end{equation}
capturing how many standard deviations the empirical fractions of links deviate from the expectation if collaborations would not be affected by gender preferences.
Fig.~4\textbf{b} in the main text thus show the evolution of  $z_{ff}$ and $z_{mm}$ from 1970 to 2020.

\clearpage

\section*{Career length and number of collaborators}
In the main text, we showed how the gender gap in the number of collaborators drives the emergence of unfairness in the acquisition and diffusion of scientific knowledge. 
Differences in the network size of women compared to men may stem from multiple factors. 
In particular, women are more likely to drop out of academia at all career stages compared to men~\textit{(6)}, limiting their chances of growing their network of collaborators.
Therefore, we now explore the relationship between career length, network size and gender. 

\begin{figure*}[t!]
    \centering
    \includegraphics[width=\textwidth]{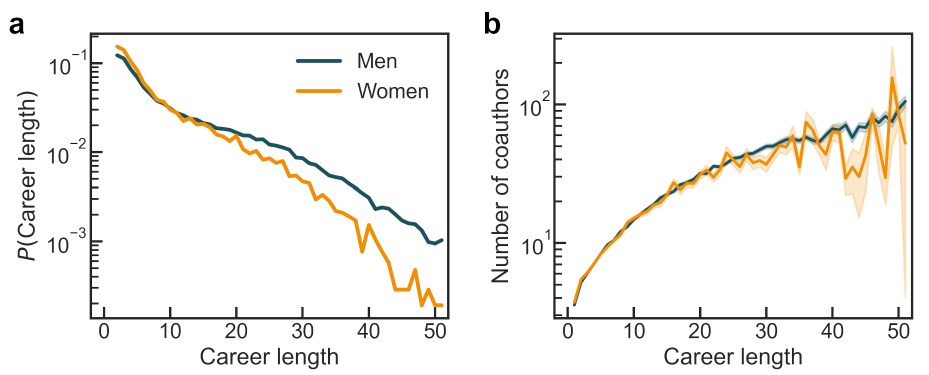}
    \caption{\textbf{Career length and number of collaborators.} Panel \textbf{a} shows the distributions of career lengths for women (yellow) and men (blue). Panel \textbf{b} displays the total number of coauthors as a function of the career length for women and men.}
    \label{fig:s4}
\end{figure*}

\cref{fig:s4}\textbf{a} shows the distributions of career lengths for women (yellow) and men (blue).
Here, the career length of an author is measured as the number of years from the first publications to the last one, and we are not showing authors with a career of one year, namely authors having first and last publications in the same year. 
We observe that women generally have shorter careers than men, confirming a higher probability of academic dropout. 
\cref{fig:s4}\textbf{b} shows that the total number of coauthors grows with the length of the career for women and men.

Together, these results suggest that career length may be one of the factors that negatively affects how women acquire and diffuse knowledge.
In fact, having shorter careers prevents women from expanding their collaboration network as much as men, which in turns limits their opportunities to acquire and diffuse scientific knowledge.

\clearpage

\let\oldthebibliography=\thebibliography
\let\oldendthebibliography=\endthebibliography
\renewenvironment{thebibliography}[1]{
    \oldthebibliography{#1}
    \setcounter{enumiv}{61}                        
}{\oldendthebibliography}

\bibliographystyle{ScienceAdvances}

\bibliography{biblio}